\documentclass[10pt,b5paper,twoside,onecolumn]{scrartcl}
\usepackage[cp1250]{inputenc}
\usepackage{color,fancyvrb}
\usepackage{graphicx}
\usepackage{hyperref}
\usepackage{fancyhdr}
\usepackage{setspace}
\usepackage{float}
\usepackage{pdfpages}
\usepackage{verbatim}
\usepackage{lmodern} 
\usepackage{amsthm}
\usepackage{amssymb}
\usepackage{multicol}
\usepackage{mathpazo}
\usepackage[T1]{fontenc}
\usepackage[titles]{tocloft}
\begin{document}
%\renewcommand{\sfdefault}{DejaVuSansCondensed}
	
% general fancy header style
\pagestyle{fancyplain}
\fancyhf{}
\fancyhead[LE]{\textit{An Analysis of a Simple Local Search Algorithm for Graph Colouring}}
\fancyhead[RO]{\textit{D. Chalupa}}
\fancyfoot[C]{\thepage}
\fancypagestyle{plain}
{
	\fancyhf{} % remove everything
	\renewcommand{\headrulewidth}{0pt} % remove lines as well
	\renewcommand{\footrulewidth}{0pt}
}

% title page	
\thispagestyle{empty}
	
\begin{center}\textbf{\LARGE\sffamily\noindent
An Analysis of a Simple Local Search Algorithm for Graph Colouring
}\end{center}

\begin{center}{\large\sffamily\noindent David Chalupa}\end{center}
\begin{center}
{
\noindent
Computer Science\\
School of Engineering and Computer Science\\
University of Hull\\
Cottingham Road\\
Hull HU6 7RX, United Kingdom\\
Tel.: +441482463069\\
Email: \texttt{D.Chalupa@hull.ac.uk}
} \vspace{15pt}\\
\textit{Acknowledgement:} This is a preprint version of an article, which is currently in review. This text will be substituted later with an acknowledgement and a link to the final version of the article.
\end{center}

\vspace{15pt}
	
\paragraph{Abstract.} Vertex Descent is a local search algorithm which forms the basis of a wide spectrum of tabu search, simulated annealing and hybrid evolutionary algorithms for graph colouring.
These algorithms are usually treated as experimental and provide strong results on established benchmarks.
As a step towards studying these heuristics analytically, an analysis of the behaviour of Vertex Descent is provided.
It is shown that Vertex Descent is able to find feasible colourings for several types of instances in expected polynomial time. This includes $2$-colouring of paths and $3$-colouring of graphs with maximum degree $3$. The same also holds for $3$-colouring of a subset of $3$-colourable graphs with maximum degree $4$. As a consequence, Vertex Descent finds a $3$-colouring in expected polynomial time for the smallest graph for which Br\'{e}laz's heuristic DSATUR needs $4$ colours. On the other hand, Vertex Descent may fail for forests with maximum degree $3$ with high probability.


\section{Introduction}

\textit{Local search} forms the basis of some of the most popular metaheuristics \cite{hoosstutzlesls}, including tabu search \cite{glover1,glover2,tabusearch} and simulated annealing \cite{simulatedannealingcerny,simulatedannealing}. These algorithms find their applications in a large spectrum of combinatorial optimisation problems. One of them, the \textit{graph colouring problem} (GCP), is among the most widely studied problems in a body of experimental and theoretical literature. It is one of the most well-known NP-hard problems \cite{introalg,karp} with applications in \textit{multiprocessor task scheduling} \cite{multiprocessor}, \textit{timetabling} \cite{timetabling} or \textit{frequency assignment in mobile radio networks} \cite{frequency}.

Formally, let $G = [V,E]$ be an undirected graph. Then, the \textit{aim of GCP} is to find $k$ disjoint colour classes $V_1$, $V_2$, ..., $V_k$ for which $\cup_{i=1}^{k} V_i = V$ and $\forall ~ \{v,w\} \in E ~ [\exists ~ i,j ~ (v \in V_i \wedge w \in V_j \wedge i \neq j)]$. A \textit{conflict} is defined as a pair of equally coloured adjacent vertices. A colouring with no conflicts is called \textit{feasible}. Otherwise it is \textit{infeasible}. Minimum $k$ for which there is a feasible colouring is called \textit{chromatic number} and is denoted by $\chi$.

Vertex Descent is a local search algorithm which represents the basic component of a large number of successful graph colouring heuristics. It tackles GCP by searching in the space of infeasible colourings and by minimising the number of conflicts between equally coloured vertices. This approach, sometimes called \textit{k-fixed penalty strategy}, is one of the most widespread and successful approaches to solve GCP \cite{localsearch}.

Algorithms which use Vertex Descent as a basis include the tabu search algorithm TabuCol \cite{tabucol}, learning-based local search algorithms \cite{porumbel::cartography}, the simulated annealing graph colouring algorithm \cite{johnson::annealing}, the quantum annealing algorithm QACol \cite{quantumann,qatuning,distributedquantum} and a number of hybrid evolutionary algorithms that use local search (mostly TabuCol) as an intensification subroutine \cite{dorne::genetic,hybrid,ie2col,lu::memetic,headcoloring,porumbel::evocoljour,extracol}.

While these are successful experimental algorithms, their theoretical understanding still seems to be very limited. On the other hand, analysis of randomised search heuristics has progressed rapidly over the last years \cite{searchheurtheory,eacomplexity}. In this paper, we provide the first analytical results on the behaviour of Vertex Descent. To the best of our knowledge, this is the first analytical study of a graph colouring heuristic based on the widely used k-fixed penalty strategy.

\paragraph{Contributions}
We show that the behaviour of Vertex Descent may be modelled using the fitness levels method \cite{lehrelevels,eacomplexity,sudholtlower} and fair random walks \cite{covertime} can be used to model its behaviour on plateaus.

For Vertex Descent, we obtain that it is able to find feasible colourings in expected polynomial time for several types of instances. These instances include:

\begin{itemize}
\item $(\Delta+1)$-colouring for graphs with maximum degree $\Delta$;
\item $\Delta$-colouring for graphs with maximum degree $\Delta$ which are neither complete graphs $K_n$ nor odd rings and;
\item $3$-colourings for $3$-colourable graphs with maximum degree $4$ such that the neighbours of each vertex with degree $4$ induce $K_1 \cup P_3$, i.e. an isolated vertex and a path of length $3$.
\end{itemize}

\noindent
The last of the instance types above includes the smallest hard-to-colour graph for Br\'{e}laz's heuristic DSATUR \cite{smallhardbrelaz}. On the other hand, we also obtain results on instances which are hard-to-colour for Vertex Descent. For forests with maximum degree $3$, Vertex Descent with $k = 2$ may not produce a feasible $2$-colouring with probability $1 - o(1)$. We also present a connected $3$-colourable graph for which Vertex Descent with $k = 3$ will not be able to find a feasible $3$-colouring with probability $1 - o(1)$.

The paper is structured as follows. In Section 2 we present an overview of the related work. In Section 3 we introduce the Vertex Descent algorithm. In Section 4 we present the general analytical results. The main results for Vertex Descent are presented in Section 5. Finally, in Section 6 we provide the conclusions and identify the problems which remain open.

\section{Related Work}

% Mention the other approches to solve graph colouring
% Move on to the analysis of local search and EAs
% Mention all the related problems and analyses

Most of the modern graph colouring heuristics are \textit{hybrid algorithms}. Currently, the most successful algorithms include a \textit{distributed quantum annealing algorithm} QACol \cite{quantumann,qatuning}, an algorithm IE$^2$Col, based on \textit{extraction and expansion} of independent sets \cite{ie2col} and a \textit{memetic algorithm} HEAD with a very small population of two colourings \cite{headcoloring}. All three of these algorithms use Vertex Descent combined with several different ideas. These include solution populations, tabu lists, thermal and quantum fluctuations, partition crossovers and preprocessing. However, Vertex Descent remains at the core of all these approaches.

One of the most well-known hybrid evolutionary algorithms for GCP was introduced by Galinier and Hao \cite{hybrid}. It used the tabu search algorithm TabuCol \cite{tabucol} as an intensification subroutine. TabuCol practically represents the Vertex Descent algorithm enhanced by a tabu list to prevent it from cycling. Glass and Pr\"{u}gel-Bennett investigated an adaptation of the hybrid algorithm by Galinier and Hao obtained by substituting TabuCol with Vertex Descent \cite{hybridanalysis}. They concluded that this version can perform comparably to the original variant but requires a larger population. Vertex Descent has also been used as a technique of search space sampling in an evaluation of different objective functions for the problem \cite{evaluationfunctionscoloring}.

This indicates that an analysis of Vertex Descent is highly relevant for a better understanding of these algorithms. Most of the theoretical results were previously proven for constructive graph colouring heuristics. These results usually focus on the number of colours used in the worst case \cite{worstcasegreedy,worstcasebrelaz}. The most popular constructive algorithms for GCP include Br\'{e}laz's heuristic \linebreak DSATUR \cite{Brelaz} and the recursive largest first heuristic \cite{leighton::scheduling}. Br\'{e}laz's heuristic is known to use $n$ colours for a $3$-colourable graph on $\mathcal{O}(n)$ vertices if an unfavourable sequence of vertices is chosen \cite{worstcasebrelaz}. There is also a $3$-colourable graph on $8$ vertices for which Br\'{e}laz's heuristic will always use $4$ colours \cite{smallhardbrelaz}.

Sudholt and Zarges \cite{ilscoloringanalysis} analysed an iterated local search algorithm for graph colouring. This algorithm operated in the space of feasible colourings and used moves called Kempe chains and colour eliminations to minimise the number of colours used. This study provided several interesting insights into the behaviour of local search for graph colouring. However, the most popular experimental approach based on the $k$-fixed penalty strategy still seems to be overlooked in the theoretical studies.

It is worth noting that an alternative $k$-fixed approach is based on searching in the space of partial feasible colourings and minimizing the number of uncoloured vertices \cite{blochliger::partialcol,morgenstern::dncs}. These approaches can also be combined \cite{vss}. Order-based and distributed algorithms have also been studied \cite{culberson,morgenstern::dncs}.

Numerous analytical results have also been obtained for other combinatorial optimisation problems. For the maximum matching problem, behaviours of local search and evolutionary algorithms for paths and worst-case approximations were investigated \cite{maxmatchanalysis}. For the vertex cover problem, behaviour of hybrid algorithms was analysed for specific types of graphs \cite{vertexcoverhybrid} and an iterated local search algorithm was investigated in the context of sparse random graphs \cite{vertexcoverrandom}. Fixed-parameter evolutionary algorithms were also analysed \cite{vertexcoverfpt}. Other studied problems include makespan scheduling \cite{makespandiscrep}, the Euclidean travelling salesperson problem \cite{tspanalysis}, the Eulerian cycle problem \cite{eulerian}, the minimum spanning tree problem \cite{minimumspanning} or the minimum cut problem \cite{minimumcuts}.

\section{The Vertex Descent Algorithm}

In this section, we briefly review the Vertex Descent algorithm. Algorithm 1 presents the pseudocode. As an input, we have a graph $G = [V,E]$ and the number of colours $k$. The output is the best colouring found in the search process.

In step 1 the vertices are assigned uniformly random colours to create an initial candidate solution $S$. In step 2 the currently best solution $S_{best}$ and its objective value $f_{best}$ are initialised. Let $s(v)$ be the colour of $v$ in $S$. Then we have that $f(S) = |\left\{\{v,w\} \in E: s(v) = s(w)\right\}|$.

\begin{table}
\begin{center}
Algorithm 1. Vertex Descent local search algorithm for graph colouring\vspace{5pt}\\
\begin{tabular}{l|l}
& Input: graph $G = [V,E]$, the number of colours $k$\\
& Output: solution $S = \{V_1,V_2,...,V_k\}$\\\hline
1  & $\forall ~ v \in V$ let $v$ have a random colour in $S$\\
2  & $S_{best} = S$, $f_{best} = f(S)$\\
3  & while stopping criteria are not met\\
4  & ~~~~let $N(S)$ be the neighbourhood of $S$\\
5  & ~~~~$[v,c] = \arg\min_{[v,c], S' \in N(S)} f(S')$\\
6  & ~~~~let $S$ be constructed by applying move $[v,c]$ to $S$\\
7  & ~~~~if $f_{best} > f(S_{best})$ then $f_{best} = f(S')$, $S_{best} = S$\\
8  & return $S_{best}$\\
\end{tabular}
\end{center}
\end{table}

An iterative procedure follows. In step 4 we scan the neighbourhood $N(S)$ of solution $S$. The set of conflicting vertices in $S$ is $C(S) = \{v \in V: \exists ~ \{v,w\} \in E ~ [s(v) = s(w)]\}$. Then, the neighbourhood $N(S)$ is a set of all solutions obtained from $S$ by recolouring any of the vertices in $C(S)$ by any of the $k-1$ other colours which the vertex can have. Therefore, there are $|N(S)| = (k-1)|C(S)|$ possible moves. In step 5 we choose the move which leads to the lowest number of conflicts. In step 6 we recolour $v$ by its new chosen colour $c$. In step 7 the best solution $S_{best}$ and its objective function value $f_{best}$ are updated, if needed. This process is repeated until a feasible colouring is found.

It is worth mentioning that the testing of the new objective values for candidate solutions $S' \in N(S)$ requires $\mathcal{O}(1)$ time if Vertex Descent is implemented using an auxiliary matrix. Therefore, step 5 requires $\mathcal{O}(kn)$ time in the worst case, since there are always at most $n$ conflicting vertices and $k-1$ possible colours.

\section{General Results}

At this point, we begin with our general analytical results. In the following, Lemma 1 and Lemma 2 provide results on the total number of conflicts and its reduction by recolouring a vertex. Next, Theorem 1 provides the general result on the number of conflicts reached by Vertex Descent in expected polynomial time for an arbitrary graph.

\paragraph{Definition 1.} Let $S$ be a colouring of a graph $G = [V,E]$ and let $s(v)$ be the colour of $v$ in $S$. Then, the function
\begin{equation}
\Gamma_S(v,c) = \left|\{ v' \in V: \{v,v'\} \in E \wedge s(v') =  c\}\right|
\end{equation}
denotes the number of neighbours of a vertex $v$ with colour $c$. 

\paragraph{Lemma 1.} Let $S$ be a colouring of an arbitrary graph $G = [V,E]$ and let $s(v)$ be the colour of $v$ in $S$. Let $C(S)$ be the set of conflicting vertices in $S$. Then, the total number of conflicts $confl(S)$ in $S$ is:
\begin{equation}
confl(S) = \frac{1}{2} \sum_{v \in V} \Gamma_S(v, s(v)) = \frac{1}{2} \sum_{v \in C(S)} \Gamma_S(v, s(v)).
\end{equation}

\paragraph{Proof.} Consider the subgraph $G'$ induced by the set of conflicting vertices $C(S)$, containing only edges between these vertices. In such a subgraph, each conflicting vertex $v$ has degree $\Gamma(v, s(v))$. We obtain that $\sum_{v \in C(S)} deg_{G'}(v) = \sum_{v \in C(S)} \Gamma_S(v, s(v)) = 2 confl(S)$. Add the non-conflicting vertices of $G$ as isolated vertices to our induced subgraph $G'$ to form a new graph $G''$. We now have that $\sum_{v \in V} deg_{G''}(v) = \sum_{v \in V} \Gamma_S(v, s(v)) = 2 confl(S)$. 
$\blacksquare$

\paragraph{Lemma 2.} Let $G = [V,E]$ be an arbitrary graph a let $S$ be a $k$-colouring of $G$. If there is a vertex $v$ such that $\Gamma_S(v,c) > \left\lfloor\frac{deg(v)}{k}\right\rfloor$, then there is a move recolouring $v$ by colour $c'$ such that $\Gamma_S(v,c') \leq \left\lfloor\frac{deg(v)}{k}\right\rfloor$. Additionally, it is a move which always leads to a drop in the number of conflicts.

\paragraph{Proof.} Consider the move recolouring $v$ with colour $c'$ in a colouring $S$ such that $\Gamma_S(v,c')$ is minimal. After such a move, the number of conflicts cannot become higher. Hence, the move will be accepted. We have that $\Gamma_S(v,c') \leq \left\lfloor\frac{deg(v)}{k}\right\rfloor$. This is implied by the fact that the worst case involves evenly distributed colours between the neighbours of $v$. Since $\Gamma_S(v,c') < \Gamma_S(v,c)$, the move leads to a drop in the number of conflicts by at least one.
$\blacksquare$

\vspace{10pt}\noindent
To analyse the behaviour of Vertex Descent in the phase of decreasing number of conflicts, we will use the method of \textit{fitness levels} \cite{lehrelevels,eacomplexity,sudholtlower}. In this method, the search space is partitioned into levels such that each level contains colourings with an equal number of conflicts. Then, the expected time to obtain a colouring with a certain number of conflicts is derived as a sum of expected waiting times for improvements from the current fitness level to a better one. Theorem 1 provides a result on the fitness level reached by Vertex Descent in expected polynomial time.

\paragraph{Theorem 1.} Let $G = [V,E]$ be an arbitrary graph on $n$ vertices and $m$ edges. Then, after $\mathcal{O}(knm)$ time in expectation, Vertex Descent for an instance of GCP with $k$ colours will arrive at a solution $S$ with each vertex $v$ involved in at most $\left\lfloor\frac{deg(v)}{k}\right\rfloor$ conflicts. The total number of conflicts in $S$ will be:
\begin{equation}
confl(S) \leq \frac{1}{2} \sum_{v \in V} \left\lfloor\frac{deg(v)}{k}\right\rfloor.
\end{equation}

\paragraph{Proof.} Let us have a $k$-colouring $S$ of $G$. If there is a vertex $v$ involved in more than $\left\lfloor\frac{deg(v)}{k}\right\rfloor$ conflicts in $S$, then Lemma 2 implies that $v$ can be recoloured to obtain an improvement by at least one conflict. We now investigate the expected time until Vertex Descent arrives at a colouring for which each vertex is involved in at most $\left\lfloor\frac{deg(v)}{k}\right\rfloor$ conflicts. From Lemma 1, we have that there are at most $\frac{1}{2} \sum_{v \in V} \left\lfloor\frac{deg(v)}{k}\right\rfloor$ conflicts in such a colouring.

Let us pessimistically assume that the initial solution contains the maximum number of conflicts possible which is $m$. Then, for each move, we have that Vertex Descent obtains an improvement by at least one conflict. Hence, the number of moves needed to obtain our target solution is:

\begin{equation}
m - \frac{1}{2} \sum_{v \in V} \left\lfloor\frac{deg(v)}{k}\right\rfloor \leq m.
\end{equation}

\noindent
This simplified upper bound already allows us to see that $\mathcal{O}(m)$ moves are needed, with $\mathcal{O}(kn)$ being the cost of a neighbourhood scan.
$\blacksquare$

\vspace{10pt}\noindent
Theorem 1 will further be used to upper bound the number of conflicts obtained in the initial phase of the search. However, it also implies that Vertex Descent finds $(\Delta+1)$-colourings for graphs with maximum degree $\Delta$ in expected polynomial time, including $3$-colourings for odd rings.

\paragraph{Corollary 1.} For a graph on $n$ vertices with maximum degree $\Delta$ and $k = \Delta+1$, the expected time for Vertex Descent to find a feasible $(\Delta+1)$-colouring
is upper bounded by $\mathcal{O}(\Delta nm)$.

\paragraph{Corollary 2.} For an odd ring on $n$ vertices and $k = 3$, the expected time for Vertex Descent to find a feasible $3$-colouring
is upper bounded by $\mathcal{O}(n^2)$.

\section{Main Results for Vertex Descent}

Theorem 1 can be used to analyse the behaviour of Vertex Descent for certain types of graphs. However, in cannot be used directly if the algorithm needs to perform an exploration of a plateau, i.e. a set of solutions with an equal number of conflicts. This is the first scenario we consider in our further investigations.

Next, we focus on $3$-colourable graphs with maximum degree $4$ for which the neighbours of all vertices with degree $4$ induce $K_1 \cup P_3$. For these graphs, we show that Vertex Descent finds a feasible $3$-colouring in expected polynomial time.

In the last part of the analysis, we present a forest with maximum degree $3$ for which Vertex Descent does not produce a feasible $2$-colouring with probability $1-o(1)$. This idea is then generalised to connected $3$-colourable graphs. We find a $3$-colourable graph for which Vertex Descent does not find a feasible $3$-colouring with probability $1-o(1)$.

\subsection{$\Delta$-colouring of Graphs with Maximum Degree $\Delta$ Other than $K_n$ and Odd Rings}

Vertex Descent often has to search on plateaus of colourings with an equal number of conflicts. The most straightforward example for this is represented by $2$-colouring of paths for which the plateaus are formed by conflicts located at different positions on the path. Exploration of these plateaus by Vertex Descent can be modelled as a random walk of a conflict on the path.

Theorem 2 provides a more general result. For graphs with maximum degree $\Delta$ which are neither complete graphs $K_n$ nor odd rings, we show that Vertex Descent will produce a feasible $\Delta$-colouring in expected polynomial time.

It is also worth noting that the upper bounds in the following results are based on rather pessimistic assumptions. The aim is to show that Vertex Descent solves the instance in expected polynomial time. In practice, the algorithm usually tends to be much faster.

\begin{figure}
\begin{center}
\includegraphics[scale=1.25]{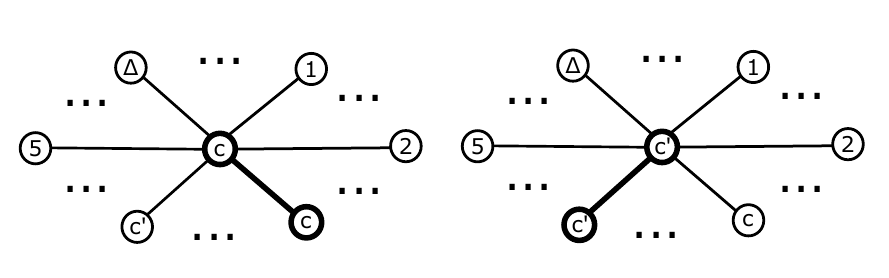}
\end{center}
\caption{An illustration of a local conflict in $\Delta$-colouring of a graph $G$ with maximum degree $\Delta$ which is neither $K_n$ nor an odd ring. After the first phase of Vertex Descent, each vertex is involved in at most one conflict. Recolouring of $v$ from $c$ to $c'$ reveals that for this type of graphs, behaviour of Vertex Descent can be modelled using fair random walks.}
\end{figure}

\paragraph{Theorem 2.} Let $G$ be a connected graph on $n$ vertices and $m$ edges with maximum degree $\Delta$ which is neither $K_n$ nor an odd ring. Then, the expected time for Vertex Descent with $k = \Delta$ to find a feasible $\Delta$-colouring for $G$ is upper bounded by $\mathcal{O}(\Delta n^4m)$.

\paragraph{Proof.} Brooks' theorem states that for a connected graph on $n$ vertices with maximum degree $\Delta$, it holds that it is $\Delta$-colourable if and only if it is neither $K_n$ nor an odd ring \cite{Brooks}. This already establishes that there is a feasible $\Delta$-colouring of $G$.

Based on Theorem 1, we have that after $\mathcal{O}(\Delta nm)$ time in expectation, vertices with degree at most $\Delta-1$ will be involved in no conflicts and vertices with degree $\Delta$ will be involved in at most one conflict. Figure 1 illustrates an example of such a vertex $v$ (in the middle) with degree $\Delta$. All $\Delta$ colours must be used between the neighbours of vertex $v$, since otherwise one colour would be left for use and the conflict could be resolved in one step.

In Vertex Descent, the choice of a new colour $c'$ for $v$ is uniformly distributed, unless one of the possible colours leads to a resolution of the conflict. There are $\Delta-1$ possible choices of this new colour, each occurring with probability $1 / (\Delta-1)$. Note that there are two conflicting vertices for each conflict which must have degree $\Delta$ and both of them must be involved only in this one conflict.

After recolouring $v$ from $c$ to $c'$, the conflict ``moves'' in the sense that it gets resolved and possibly replaced by a conflict with another neighbour of $v$, as illustrated by Figure 1. This process leads to a random walk of each of these conflicting vertices. This random walk is fair which is implied by the probability arguments described above.

The cover time of such a random walk is $\mathcal{O}(nm)$. After this number of steps of Vertex Descent affecting a fixed conflict in expectation, this conflict will visit a sequence of positions such that each vertex was involved in the conflict at least once. Since $G$ is $\Delta$-colourable, there must be a vertex $v$ such that if $v$ becomes conflicting in a colouring $S$, then there is a colour $c$ such that $\Gamma_S(v,c) = 0$. If there was no such vertex, then for all vertices all $\Delta$ colours would be used to colour their neighbours, leading to $G$ not being $\Delta$-colourable. When such a vertex becomes conflicting, the conflict can be resolved in one step of Vertex Descent.

For each of the $\mathcal{O}(n)$ remaining fitness levels, we now consider one specific conflict and assume that other moves will not lead to an improvement. In each fitness level, it takes $\mathcal{O}(n)$ neighbourhood scans in expectation to pick a move affecting our fixed conflict and $\mathcal{O}(\Delta n)$ is the complexity of such a neighbourhood scan. The cover time of the random walk analysed above is $\mathcal{O}(nm)$ which proves the theorem.
$\blacksquare$

\vspace{10pt}\noindent
The previous result implies that Vertex Descent finds feasible $2$-colourings for paths and even rings and feasible $3$-colourings for graphs with maximum degree $3$ in expected polynomial time.

\paragraph{Corollary 3.} For both paths and even rings on $n$ vertices and $k = 2$, the expected time for Vertex Descent to find a feasible $2$-colouring
is upper bounded by $\mathcal{O}(n^5)$.

\paragraph{Corollary 4.} Let $G$ be a 3-colourable graph on $n$ vertices and $m$ edges a let its maximum degree be $\Delta \leq 3$. Then, the expected time for Vertex Descent with $k = 3$ to find a feasible $3$-colouring for $G$ is upper bounded by $\mathcal{O}(n^4m)$.

\subsection{$3$-colouring of a Subset of $3$-colourable Graphs with Maximum Degree $4$}

Kochol et al. have explored the $3$-colorability problem for $3$-colorable graphs with maximum degree $4$ \cite{colorability3maxdegree4}. They have shown that these instances can be partitioned into easy and hard, depending on the subgraphs induced by the neighbours of vertices with degree $4$. One of the patterns leading to solvability in polynomial time is when the neighbours of each vertex with degree $4$ induce $K_1 \cup P_3$, i.e. a subgraph with an isolated vertex and a path on $3$ vertices. One of their results was that a $3$-colorability problem for a graph which only contains patterns leading to polynomial-time solvability can be transformed into a $3$-colorability problem for a graph with neighbours of each vertex with degree $4$ inducing $K_1 \cup P_3$.

In the following, we show that Vertex Descent will find a feasible $3$-colouring for such a graph in expected polynomial time.

\paragraph{Theorem 3.} Let $G$ be a 3-colourable graph on $n$ vertices and $m$ edges a let its maximum degree be $\Delta \leq 4$. Furthermore, let the neighbours of each vertex with degree $4$ induce $K_1 \cup P_3$ as a subgraph. Then, the expected time for Vertex Descent with $k = 3$ to find a feasible $3$-colouring for $G$ is upper bounded by $\mathcal{O}(n^4m)$.

\paragraph{Proof.} Based on Theorem 1, we have that after $\mathcal{O}(nm)$ time in expectation, vertices with degree at most $2$ will be involved in no conflicts and vertices with degree $3$ or $4$ will be involved in at most one conflict. For vertices with degree $3$, the arguments used in the proof of Theorem 2 can be applied to show that conflicting vertices with degree $3$ perform a fair random walk on the graph.

\begin{figure}
\begin{center}
\includegraphics[scale=1.5]{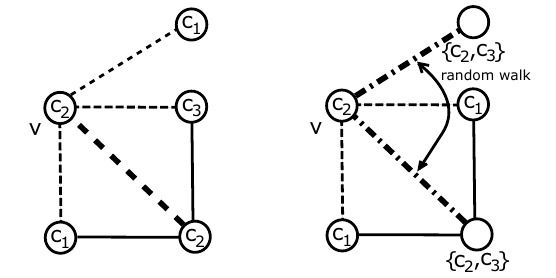}
\end{center}
\caption{A local illustration of a vertex $v$ with degree $4$ and its neighbours which induce subgraph $K_1 \cup P_3$, investigated in Theorem 3. By using enumeration, it can be shown that only two scenarios are of a particular interest in this subgraph. In the scenario on the left-hand side, Case 2a for Theorem 3 is depicted. Vertex $v$ will be recoloured from $c_2$ to $c_3$, effectively moving the conflict to a vertex with degree $2$ which can be resolved in one step. On the right-hand side, we have a scenario corresponding to Case 2b. This case involves a random walk of a conflict, since both $c_2$ and $c_3$ can be used to colour the respective vertices.}
\end{figure}

Consider a vertex $v$ with degree $4$ which is involved in a conflict. Figure 2 illustrates a vertex $v$ and its neighbours in $G$ and possible scenarios in the search. Note that the neighbours of our vertex $v$ with degree $4$ induce the subgraph $K_1 \cup P_3$, with edges drawn in full lines. The other edges are drawn using dashed lines. There are 
$4$ neighbours of $v$ and $3$ colours. Therefore, one colour must be used twice between the neighbours of $v$ and the other two only once.

Let $c_1$ be the colour which is used twice. This implies that $v$ must be coloured by $c_2$ or $c_3$. Without loss of generality, fix $c_2$ as the current colour of $v$. Let the vertices with degree $2$ in Figure 2 be called \textit{pick vertices} and let them be denoted by $v_p^1$ and $v_p^2$. Furthermore, let $c(v)$ denote the current colour of $v$. We now use enumeration based on colours of the pick vertices.

\textit{Case 1.}  Let $c(v) = c(v_p^1) \vee c(v) = c(v_p^2)$. This implies that $v$ is in a conflict with a vertex with degree $2$. This conflict can be resolved in one step of Vertex Descent.

\newpage
\textit{Case 2.}  Let $c(v) \neq c(v_p^1) \wedge c(v) \neq c(v_p^2)$. In this situation, we have to further consider whether $v_p^1$ and $v_p^2$ are coloured differently or equally.

\textit{Case 2a.}  Let $c(v_p^1) \neq c(v_p^2)$. One of the pick vertices must be coloured by $c_1$ and the other one by $c_3$. Let the vertex in the bottom right corner be denoted by $v_o$. Then, $v_o$ cannot be coloured by $c_3$, since that colour can be used only once among the neighbours of $v$.

If $v_o$ was coloured by $c_1$, it would mean that the pick vertex coloured by $c_1$ can be recoloured by $c_3$. Since the uppermost neighbour will be coloured $c_2$, this scenario leads to Case 2b (with the role of $c_1$ and $c_3$ being switched).

Let $v_o$ be coloured by $c_2$. Consequently, the uppermost neighbour of $v$ must be coloured by $c_1$, as shown in Figure 2 (on the left-hand side). This situation leads Vertex Descent to choose $c_3$ as the next colour for $v$ which leads to Case 1 and a resolution of the conflict in one step of Vertex Descent.

\textit{Case 2b.}  Let $c(v_p^1) = c(v_p^2)$. The pick vertices must be coloured by $c_1$ in this case. Figure 3 illustrates that this situation leads to a free choice of $c_2$ and $c_3$ among the other two neighbours of $v$. Hence, we are facing a fair random walk of the conflicting vertices on the subgraph obtained by pruning the pick vertices. The cover time argument can now be applied to this situation.

Let us now assume that Case 2b happens at all times when a conflict visits a position involving a vertex with degree $4$, since the other scenarios lead to a resolution of the conflict in $\mathcal{O}(1)$ steps. Since the graph obtained by pruning all pick vertices has maximum degree at most $3$, we can now use the same arguments with cover time of random walks as in Theorem 2 which concludes the proof.
$\blacksquare$

\vspace{10pt}\noindent
At this point, it is interesting to confront this result to a result obtained by a constructive algorithm. Figure 3 illustrates graph $G_1$ which is the smallest hard-to-colour graph for Br\'{e}laz's heuristic DSATUR \cite{smallhardbrelaz}. It will use $4$ colours to colour $G_1$ for all of its possible randomised runs. On the other hand, Theorem 3 implies that Vertex Descent with $k = 3$ will find a feasible $3$-colouring for $G_1$ in expected polynomial time. Therefore, it is possible to find graphs for which Vertex Descent will outperform a well-known constructive algorithm.

\paragraph{Corollary 5.} For graph $G_1$, the expected time for Vertex Descent with $k = 3$ to obtain a feasible $3$-colouring is upper bounded by $\mathcal{O}(1)$.

\begin{figure}
\begin{center}
\includegraphics[scale=0.8]{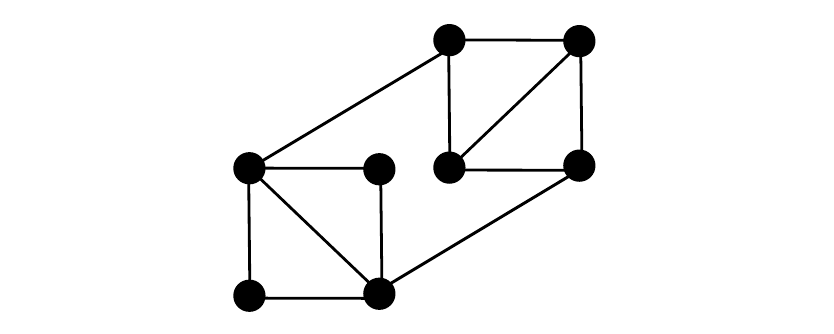}
\end{center}
\caption{An illustration of the smallest hard-to-colour graph for Br\'{e}laz's heuristic. We denote this graph by $G_1$ and it is the smallest $3$-colourable graph, for which Br\'{e}laz's heuristic will always use $4$ colours \cite{smallhardbrelaz}.}
\end{figure}

\subsection{Hard-to-colour Instances for Vertex Descent}

The previous results provide a somewhat optimistic view on the behaviour of Vertex Descent. In this section, we focus on the limitations of this algorithm.

In Theorem 4, we prove that Vertex Descent may fail to provide a feasible $2$-colouring for a forest with maximum degree $3$ with high probability. Such a forest is depicted in Figure 4 and will be denoted by $G_{2,c}$, where $c$ represents the number of identical trees in the forest.

\paragraph{Theorem 4.} For graph $G_{2,c}$ on $n = 14c$ vertices, Vertex Descent with $k = 2$ will not produce a feasible $2$-colouring with probability lower bounded by $1-(32/31)^{-\Omega(n)}$.

\paragraph{Proof.} Let us consider a single tree of $G_{2,c}$ and its vertices $A$ and $B$. Suppose that in the initial colouring, they are equally coloured. This occurs with probability $1/2$. Then, let vertices $C$, $D$, $E$ and $F$ have a different colour than $A$ and $B$. This occurs with probability $1/16$, i.e. this initial configuration is generated with probability $1/32$.

\begin{figure}
\begin{center}
\includegraphics[scale=1]{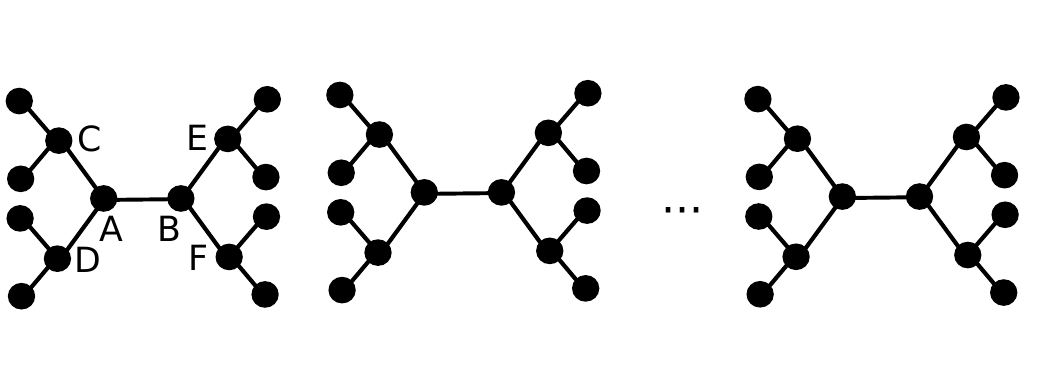}
\end{center}
\caption{An illustration of graph $G_{2,c}$ on $n = 14c$ vertices which is a forest with maximum degree $3$ and for which Vertex Descent will fail to produce a feasible $2$-colouring with high probability.}
\end{figure}

At this point, each leaf of the tree will be recoloured so that conflicts on leaves are resolved. This leaves Vertex Descent with a single conflict between $A$ and $B$. By recolouring any of these two vertices, a colouring with $2$ conflicts is reached. Next, the best move in the neighbourhood is to recolour the previous vertex back to get $1$ conflict. This leads to an infinite loop between two suboptima.

The probability that this occurs for a single tree is lower bounded by $1/32$. Therefore, the probability that this occurs for at least one tree in $G_{2,c}$ is lower bounded by $1-(31/32)^{n/14} = 1-(32/31)^{-\Omega(n)}$.
$\blacksquare$

\vspace{10pt}\noindent
This confirms that Vertex Descent may fail to produce a feasible $2$-colouring for forests with maximum degree $3$. This is a somewhat surprising disadvantage compared to Br\'{e}laz's heuristic which guarantees to construct a $2$-colouring for a bipartite graph \cite{Brelaz}.

%It is also worth mentioning that our empirical results suggest further limitations of this algorithm. For a tree $T_{5,6}$ with branching factor $5$ and depth $6$ and $k = 3$ Vertex Descent was not able to find a feasible $3$-colouring in $100$ runs limited to five seconds. A similar result was obtained for bipartite graph $G_{2,b,c}$ with $b = 20$ and $c = 100$, and $k = 4$. For this instance, we were not able to sample a feasible $4$-colouring in $100$ runs of Vertex Descent with the same time limit.

In the next analysis, we will further extend on this result and show how Vertex Descent may fail in $3$-colouring of a connected graph. Figure 5 illustrates the $3$-colourable graph $G_{3,\mathcal{L}}$. This graph consists of $\mathcal{L}$ ``legs'' for which Vertex Descent will be able to produce a colouring with one conflict in each ``leg''. However, the algorithm will then keep cycling with high probability. We formalise this result in Theorem 5.

\begin{figure}
\begin{center}
\includegraphics[scale=1.5]{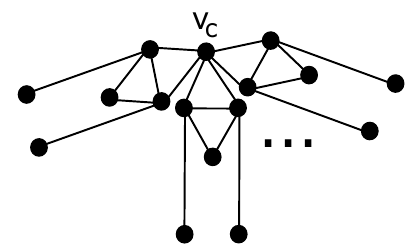}
\end{center}
\caption{An illustration of graph $G_{3,\mathcal{L}}$, consisting of $\mathcal{L}$ ``legs''. Each ``leg'' consists of a diamond and two leaves attached to it, while the central vertex is not considered a part of the ``legs''. For $3$-colouring of $G_{3,\mathcal{L}}$, Vertex Descent will tend to get stuck with a high probability.}
\end{figure}

\paragraph{Theorem 5.}
For graph $G_{3,\mathcal{L}}$ on $n = 5\mathcal{L} + 1$ vertices, Vertex Descent with $k = 3$ will not be able to find a feasible $3$-colouring with probability $1 - o(1)$.

\paragraph{Proof.}
Let the vertex of $G_{3,\mathcal{L}}$ with maximum degree be called the central vertex and let it be denoted by $v_c$. We also recall that vertices with degree $2$ will be called pick vertices and vertices with degree $1$ will be referred to as leaves.

Theorem 1 implies that after $\mathcal{O}(n^3)$ time in expectation, vertices with degree $1$ or $2$ will be involved in no conflicts, vertices with degree $4$ will be involved in at most one conflict and $v_c$ will be involved in at most $\left\lfloor\frac{2\mathcal{L}}{3}\right\rfloor$ conflicts. Within a specific leg, the neighbours of $v_c$ are adjacent to each other. Therefore, if both neighbours have the same colour as $v_c$, then they are involved in $2$ conflicts which can be reduced to $1$, according to Lemma 2. Hence, at this point, there is at most $1$ conflict per each leg.

\begin{figure}
\begin{center}
\includegraphics[scale=1.5]{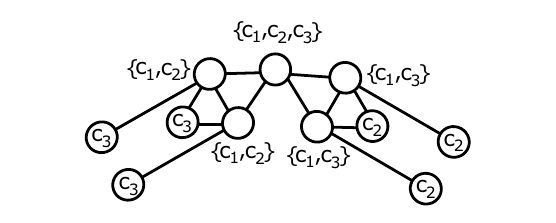}
\end{center}
\caption{An illustration of a situation in colouring of graph $G_{3,\mathcal{L}}$ which leads Vertex Descent to get stuck in an infinite loop, since colours $c_3$ or $c_2$ are ``blocked'' for vertices with degree $4$ in the two depicted legs. Without a worsening, a conflict cannot be resolved, since the central vertex would have to coloured $c_2$ and $c_3$ at the same time.}
\end{figure}

Consider the scenario illustrated by Figure 6. Let the neighbours of $v_c$ be coloured by $c_1$ or $c_2$ and let their other neighbours in the leg be coloured by $c_3$. In another leg, let the neighbours of $v_c$ be coloured by $c_1$ or $c_3$ and let their other neighbours in the leg be coloured by $c_2$.

For one of the legs, the vertices with degree $4$ cannot be recoloured by $c_3$ and for the other leg, they cannot be recoloured by $c_2$ without a worsening. This is because of the colours of pick vertices and leaves which cannot be changed, since that would require them to become conflicting. To obtain that, a worsening would have to occur first. Therefore, only the colours of vertices with degree $4$ and $v_c$ will be changed. To obtain a feasible colouring of one of the legs without recolouring of the pick vertices or leaves, $v_c$ would have to be coloured by $c_2$, and for the other leg, it would have to be coloured by $c_3$ at the same time which is in a contradiction.

\newpage
The probability of choosing $c_1$ or $c_2$ for the neighbours of $v_c$ and $c_3$ for their other three neighbours in a particular leg is at least $4 / 3^5 = 4 / 243$. The same argument holds for the case of the second leg with $c_2$ and $c_3$ being reversed. Therefore, the probability of this scenario happening in at least one leg with $c_2$ being the colour of vertices with degree $4$ and at least one other leg with $c_3$ being the colour of these vertices, is lower bounded by:

\begin{equation}
\left[ 1 - \left(\frac{239}{243}\right)^{\mathcal{L}} \right] \left[ 1 - \left(\frac{239}{243}\right)^{\mathcal{L}-1} \right] = 
1 - o(1). ~ \blacksquare
\end{equation}

\vspace{10pt}\noindent
Although $G_{2,c}$ and $G_{3,\mathcal{L}}$ can be optimally coloured by other methods, their importance lies in their hardness for Vertex Descent. Theorem 4 and Theorem 5 clearly indicate that to colour $G_{2,c}$ and $G_{3,\mathcal{L}}$ optimally, a local search algorithm must use an additional component such as thermal fluctuations or a tabu list. Behaviour of simulated annealing \cite{johnson::annealing} and tabu search \cite{tabucol} algorithms for these instances will be very interesting to investigate. Their behaviour for hard 3-colourable instances will also be interesting to explore \cite{hardthreecoloring}. We believe that our results may pave the way to better understand the interplay between different components in hybrid graph colouring heuristics and lead to theoretically underlaid parameter tuning techniques.

\section{Conclusions}

We have analysed the \textit{Vertex Descent} local search algorithm for graph colouring. The behaviour of this algorithm can be modelled using fitness levels method and its search on plateaus can be modelled using fair random walks.

It has been shown that Vertex Descent finds $(\Delta+1)$-colourings for graphs with maximum degree $\Delta$ in expected polynomial time. It also obtains feasible $\Delta$-colourings in expected polynomial time for connected graphs with maximum degree $\Delta$ which are neither complete graphs $K_n$ nor odd rings. A similar polynomial-time result has been obtained for $3$-colourable graphs with maximum degree $4$ for which neighbours of each vertex with degree $4$ induce $K_1 \cup P_3$, i.e. an isolated vertex and a path of length $3$.

However, Vertex Descent may fail for $2$-colouring of a forest with maximum degree $3$ with high probability. We have also demonstrated how Vertex Descent can get stuck in an infeasible colouring region for a $3$-colouring instance.

Vertex Descent is the basis for other local search algorithms for graph colouring, including simulated annealing \cite{johnson::annealing} and tabu search \cite{tabucol}. The most successful state-of-the-art algorithms for the problem \cite{ie2col,headcoloring,quantumann} also use Vertex Descent as one of their components.

\newpage
Based on the previous experimental studies \cite{computationalcomparisoncoloring}, it has long been known that different colouring methods work for different instances. To support a well-informed design of modern experimental graph colouring heuristics, further analytical investigations of these algorithms will be needed. Relating these analyses to experimental studies will be especially useful. We believe that our results have laid down the basis for understanding of the strengths and limitations of the approaches based on local search.

\bibliography{common}{}
\bibliographystyle{plain}

\end{document}